\documentclass[conference]{IEEEtran}
\IEEEoverridecommandlockouts

\usepackage{cite}
\usepackage{amsmath,amssymb,amsfonts}
\usepackage{algorithmic}
\usepackage{algorithm}
\usepackage{graphicx}
\usepackage{textcomp}
\usepackage{comment}
\usepackage{float} 
\usepackage{xcolor}
\usepackage{array}
\usepackage{booktabs}
\usepackage{multirow}
\usepackage{diagbox}
\usepackage{makecell}
\def\BibTeX{{\rm B\kern-.05em{\sc i\kern-.025em b}\kern-.08em
    T\kern-.1667em\lower.7ex\hbox{E}\kern-.125emX}}

\begin{document}

\title{Model Forensics in AI-Native Wireless Networks: Taxonomy, Applications, and Case Study\\}

\author{
Pengyu Chen,
Weiyang Li,
Jin Xu,
Jiacheng Wang,~\IEEEmembership{Member,~IEEE,}\\
Ning Wang,~\IEEEmembership{Member,~IEEE,}
Dusit Niyato,~\IEEEmembership{Fellow, IEEE,}
and Tao Xiang,~\IEEEmembership{Senior Member,~IEEE}
             
\thanks{Pengyu Chen, Weiyang Li, Jin Xu, Ning Wang(Corresponding Author), and Tao Xiang are with the College of Computer Science, Chongqing University, Chongqing 400044, China(e-mail: 20204227@stu.cqu.edu.cn; weiyangli@stu.cqu.edu.cn; 20232279@stu.cqu.edu.cn; nwang5@cqu.edu.cn; txiang@cqu.edu.cn). Jiacheng Wang and Dusit Niyato are with the College of Computing and Data Science, Nanyang Technological University, Singapore 639798(e-mail: jiacheng.wang@ntu.edu.sg; dniyato@ntu.edu.sg).}
}

\maketitle

\begin{abstract}
As artificial intelligence (AI) is increasingly embedded in wireless networks, models are becoming core components that influence signal processing, resource scheduling and network control.
However, model anomalies, tampering and malicious functions also introduce new security risks.
In this article, we focus on model forensics in AI-native wireless networks.
Specifically, we first discuss key problems including model authenticity verification, malicious function identification and accountability tracing, and summarize the main categories of model forensics.
We then explain the role of model forensics in AI-native wireless networks and review representative application scenarios.
In the case study, we use RF fingerprinting as an example and present two concrete workflows based on watermark authentication and backdoor detection, illustrating how provenance authentication and malicious behavior identification can be implemented in practice.
The results show that model forensics can provide important support for anomaly assessment, provenance tracing and trustworthy operation in AI-native wireless networks.
Finally, we outline several promising directions for future research in this emerging area.

\end{abstract}


\section{Introduction} \label{introduction}

As 6G continues to evolve, wireless networks are moving beyond a connectivity-focused paradigm.
They are evolving into intelligent systems shaped by AI-driven operation, task-level coordination and autonomous control\cite{1zhang2026toward}.
AI is no longer limited to offline optimization or local performance improvement.
It is becoming an integral part of wireless networks and directly participates in decision-making for signal processing, resource scheduling, network control, service management and edge collaboration\cite{cui2025overview}.
In this context, AI-native wireless networks are widely regarded as an important direction for next-generation wireless communication systems.

As models are increasingly used in performance critical control of wireless systems, potential risks are shifting from localized performance variations to system-level failure threats.
Input perturbations, backdoor injection and model capability leakage can all cause abnormal model behavior which may further appear at the system-level such as decision drift, scheduling errors or alarm failures.
As a result, abnormal model behaviors should not be regarded only as algorithmic errors.
They can also affect the trustworthy operation of AI-native wireless networks.
Systematic tracing of model anomalies helps identify the source of the problem, explain how it is triggered and support subsequent corrective actions, responsibility assessment and security auditing\cite{2motalleb2025towards}.

However, existing studies on AI security mainly focus on attack construction, robustness enhancement and defense design.
These studies are important for reducing attack impact, but their primary goal is still prevention and mitigation, rather than post-deployment investigation.
When model-related incidents have already occurred, they often cannot explain what triggered the anomaly, whether it originated from the model itself or from external factors, and whether the resulting conclusions can be reproduced and verified\cite{3senevirathna2024survey}.
This limitation motivates a forensic perspective that focuses on evidence organization, anomaly attribution, and result auditing.
In this context, model forensics provides such a perspective by examining evidence related to model identity, abnormal behavior, malicious function, capability misuse, and lifecycle traces\cite{behzadan2020founding}.

Applying this perspective to AI-native wireless networks also requires additional consideration of wireless-specific factors.
In wireless settings, model-related evidence is naturally coupled with wireless-domain observations and system-level contexts, such as channel conditions, device mobility, signal characteristics, protocol states, and lifecycle records.
As a result, the same abnormal behavior may involve both model-level causes and wireless-domain factors.
For instance, device misidentification in RF fingerprinting may be caused by model tampering, channel variation, or hidden backdoor behavior, while abnormal beam decisions may come from sensing errors, environmental changes, or limited model robustness.
Therefore, wireless model forensics requires a combination of forensic perspectives, so that authenticity, abnormal behavior, malicious function, performance degradation, and provenance can be examined jointly.

\begin{figure*}[t!]
    \centering
    \includegraphics[width=1.9\columnwidth]{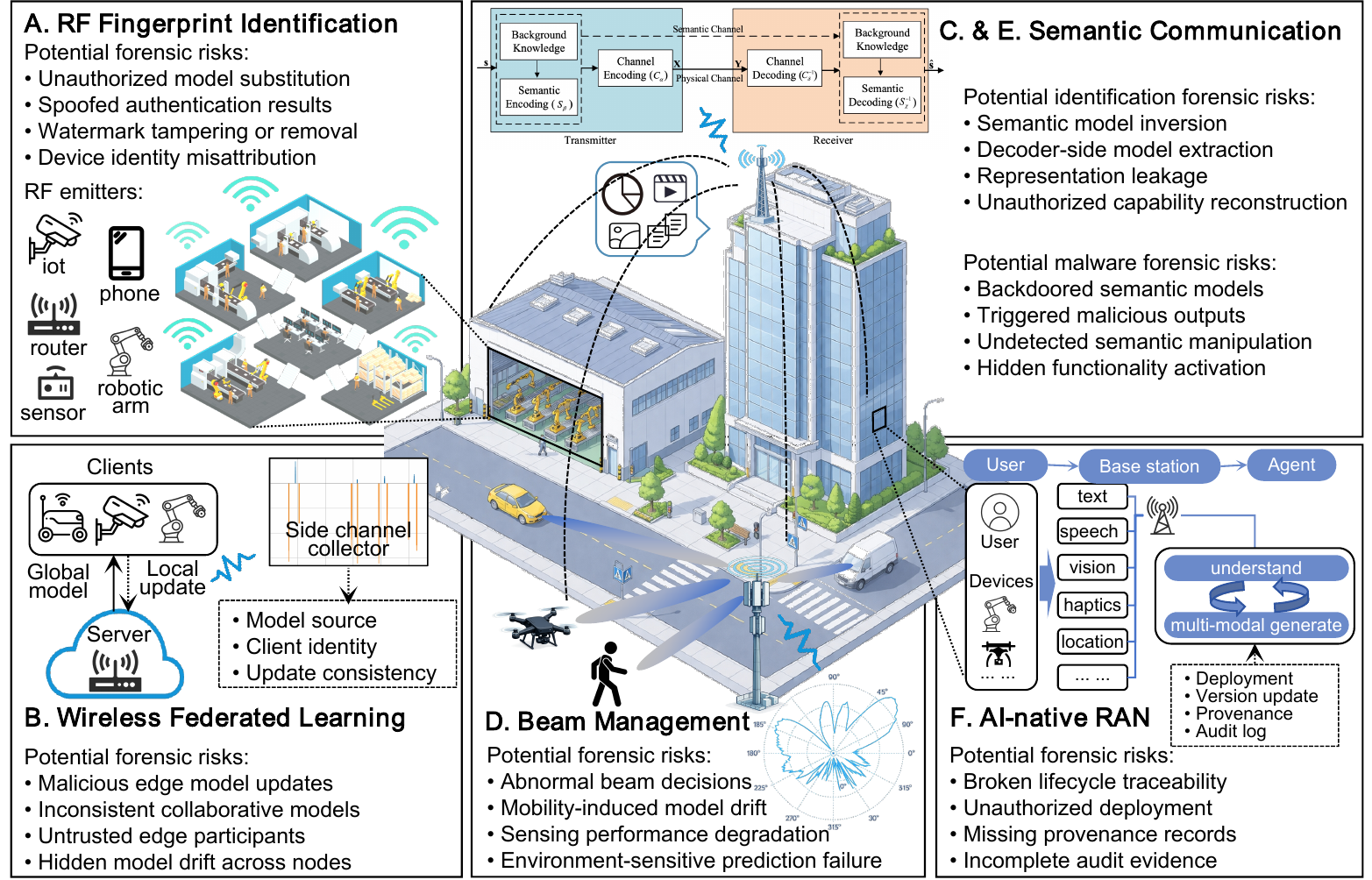}
    \caption{Representative application scenarios and potential forensic risks of AI models in AI-native wireless networks. (A) RF fingerprint identification corresponds to model authentication forensics. (B) Wireless federated learning corresponds to model fingerprinting and ballistics forensics. (C/E) Semantic communication corresponds to model identification and extraction forensics, as well as model malware forensics. (D) Beam management corresponds to model performance forensics. (F) AI-native RAN corresponds to model chain-of-custody forensics.}
    \label{fig1}
\end{figure*}

Based on these observations, we provide a systematic investigation of model forensics in AI-native wireless networks. 
First, we define model forensics and summarize its taxonomy in the context of AI-native wireless networks.
Then, we review representative application scenarios to show the roles of different forensic subareas.
Finally, we propose a unified analysis framework and present a case study to illustrate how model authentication and malware forensics can be implemented in practice.
We aim to provide a structured investigation perspective for trustworthy model operation and evidence-oriented model forensics in AI-native wireless networks.

\section{Overview of Model Forensics And AI-Native Wireless Networks} \label{overview}

This section first defines model forensics and summarizes its six main categories.
It then introduces the basic concepts of AI-native wireless networks and presents representative functional scenarios. 
Figure \ref{fig1} gives a visual overview of these representative AI-native wireless scenarios.

\subsection{Definition of Model Forensics}
Model forensics refers to forensic analysis centered on the model itself and the evidence directly related to its operation\cite{behzadan2020founding}.
Unlike security defenses that aim to prevent attacks or improve robustness, model forensics focuses on how to reach conclusions that are reproducible, auditable and attributable after anomalies, disputes or security incidents occur.
Existing studies analyze model-related evidence at multiple levels to support forensic judgments in wireless scenarios.
Representative directions include using watermarks to verify the identity and integrity of RF fingerprinting models, using hidden representations to detect backdoors in semantic communication models, and using performance degradation patterns to localize failure sources in beam prediction.
Model forensics is therefore not a single analysis task, but a set of model-centered forensic activities that address model authenticity, behavioral anomalies, capability identification, and provenance tracing in AI-native wireless networks.
The following subsection further organizes these aspects into major forensic categories.

\begin{table*}[t]
\centering
\caption{Major model forensic types, their core focuses and representative evidence, together with key characteristics and representative scenarios of AI-native wireless networks.}
\label{tab1}
\renewcommand{\arraystretch}{1.18}
\setlength{\tabcolsep}{4pt}
\setlength{\arrayrulewidth}{0.4pt}

\resizebox{\textwidth}{!}{
\begin{tabular}{
|>{\raggedright\arraybackslash}p{2.65cm}
|>{\raggedright\arraybackslash}p{4.2cm}
|>{\raggedright\arraybackslash}p{4.2cm}
|>{\raggedright\arraybackslash}m{6cm}|
}
\hline

\multicolumn{1}{|c|}{\diagbox[width=2.65cm]{\textbf{Type}}{\textbf{Properties}}}
& \multicolumn{1}{c|}{\textbf{Core Focus}}
& \multicolumn{1}{c|}{\textbf{Typical Evidence}}
& \multicolumn{1}{c|}{\textbf{AI-Native Wireless Networks}} \\

\noalign{\global\arrayrulewidth=0.25pt}
\hline
\noalign{\global\arrayrulewidth=0.4pt}

\textbf{Model Authentication Forensics}
& Verify authorized identity and integrity of the deployed model
& Watermark responses, query-response consistency, authentication outputs, and deployment integrity indicators.
& \multirow{3}{=}{\parbox[c]{6.0cm}{
\vspace{0.5em}

\textbf{a. Characteristics:}

\hspace*{1.45em}$\bullet$ Communication, computation and intelligence integration

\hspace*{1.45em}$\bullet$ Edge-cloud collaborative intelligence

\hspace*{1.45em}$\bullet$ Closed-loop and task-oriented service delivery

\hspace*{1.45em}$\bullet$ Lifecycle-aware model deployment and orchestration
}}\\
\cline{1-3}

\textbf{Model Fingerprinting and Ballistics Forensics}
& Trace model origin, vendor style, or implementation source
& Behavioral signatures, wireless side-channel traces, update-source patterns, and model-specific fingerprints.
& \\
\cline{1-3}

\textbf{Model Identification and Extraction Forensics}
& Infer model capability, architecture, or extractable function
& Query--response pairs, semantic representations, decoder outputs, and reconstruction artifacts.
& \\
\cline{1-3}

\textbf{Model Performance Forensics}
& Diagnose performance degradation and failure causes
& Prediction errors, beam decisions, sensing outputs, robustness degradation, and environment-sensitive failures.
& \multirow{3}{=}{\parbox[c]{6.0cm}{
\textbf{b. Representative Scenarios:}

\hspace*{1.45em}$\bullet$ RF fingerprint identification

\hspace*{1.45em}$\bullet$ Wireless federated learning

\hspace*{1.45em}$\bullet$ Semantic communication

\hspace*{1.45em}$\bullet$ Beam management

\hspace*{1.45em}$\bullet$ AI-native RAN lifecycle management

\hspace*{1.45em}$\bullet$ Channel estimation and signal detection

\hspace*{1.45em}$\bullet$ Resource scheduling and mobility management

\hspace*{1.45em}$\bullet$ 
... ...
}}\\
\cline{1-3}

\textbf{Model Malware Forensics}
& Reveal hidden malicious logic or trigger-dependent behavior
& Trigger responses, hidden activations, poisoned behaviors, and targeted output deviations.
& \\
\cline{1-3}

\textbf{Model Chain-of-Custody Forensics}
& Track model lifecycle, version history, and operational changes
& Version records, deployment logs, provenance chains, update history, and audit evidence.
& \\

\hline
\end{tabular}
}
\end{table*}

\subsection{Taxonomy of Model Forensics}
Forensic analysis centered on the model itself can be divided into several subcategories\cite{edwards2022exploring}: model authentication forensics, model fingerprinting and ballistics forensics, model identification and extraction forensics, model performance forensics, model malware forensics and model chain of custody.
Each subcategory has clear objectives and typical evidence types.
As illustrated in Fig.~\ref{fig1}, different AI-native wireless scenarios may show different abnormal behaviors, which lead to different forensic concerns.
\begin{itemize}
    \item \textbf{Model authentication forensics} focuses on verifying model authenticity. 
The key question is whether the model under investigation is an authorized instance and whether it has been replaced or spoofed during transfer or deployment.
Typical evidence includes model watermarks and software-level hashes, which can be used to validate the identity of wireless models, such as RF fingerprinting classifiers for device authentication.
    \item \textbf{Model fingerprinting and ballistics forensics} aims to identify clues about a model's type, origin and creator, especially when wireless models are provided by different vendors, clients, or edge participants.
It is similar to classical ballistics, where trace patterns are used to infer weapon type and provenance.
It focuses on identifying model lineage and source attribution from behavioral signatures, structural fingerprints, implementation artifacts, or wireless communication-side traces.
    \item \textbf{Model identification and extraction forensics} primarily targets closed-box models, such as semantic communication encoders or decoders.
It studies whether their behavior can be recovered through inversion, extraction, or approximate replication for further testing and analysis.
The value of these methods is not merely to copy a model, but also to provide a usable substitute model and supporting evidence for later forensic judgments.
    \item \textbf{Model performance forensics }studies why a wireless model fails and how the failure appears in network functions such as beam management.
The evidence typically comes from behavioral features such as internal state variables, activations, state-action estimates and class probability distributions.
Analysis of these features can help localize anomalous patterns and identify failure sources caused by sensing errors, environmental drift, or adversarial perturbations.
    \item \textbf{Model malware forensics} targets wireless models that contain backdoors, trigger-based policies and other malicious functions.
Its goal is to determine whether the model shows malicious behavior and to analyze the trigger conditions and internal mechanisms through reproducible experiments in a controlled environment.
    \item \textbf{Model chain of custody forensics} focuses on the model transfer records and accountability trails across training, distribution, deployment, updates and invocation in AI-native wireless systems.
It emphasizes continuous logging and traceable management of key lifecycle information, which provides a foundation for later auditing, responsibility assessment and process reconstruction.

\end{itemize}

These six categories do not need to be used at the same time for a single incident.
They can be regarded as analysis modules that can be combined as needed.
Authentication determines whether the model can be trusted.
Fingerprinting and extraction address whether the model can be identified or its behavior can be reconstructed.
Performance and malware analysis explain why the model fails and whether malicious behavior is involved.
Chain of custody provides the context needed to support investigative conclusions.
This taxonomy not only defines the scope of model forensics, but also provides a unified framework for discussing its role in AI-native wireless networks.

\subsection{AI-Native Wireless Networks}

AI-native wireless networks refer to a paradigm that embeds AI into the architecture and operational loops of wireless systems, making AI a native network capability\cite{cui2025overview}.
Unlike conventional wireless networks that mainly focus on connectivity establishment and data delivery, AI-native wireless networks integrate communications, computing, data and models into a unified set of network capabilities.
This enables the network to sense, decide, and coordinate control for specific tasks.
Under this paradigm, AI models are embedded into a range of representative wireless functions, spanning link-level processing as well as network control and autonomous operations.
For instance, in physical layer beam prediction, the network can use model predictions based on environmental information to determine transmission directions\cite{4raha2025security}.
At the network and system layers, RAN lifecycle management can provide unified control over model deployment, updates, switching, and rollback\cite{lin2024overview}.
Because of the deep integration of communications, computing, and intelligence, AI-native wireless networks require forensic support to trace model-related anomalies and provide accountable evidence for 6G and beyond security analysis\cite{alqabbani2023digital}.

In Table \ref{tab1}, we summarize the core concepts of model forensics and AI-native wireless networks. The next section discusses the applications of model forensics in AI-native wireless networks.

\section{Model Forensics in AI-Native Wireless Networks} \label{AI model}

This section reviews representative application scenarios in AI-native wireless networks and discusses the roles of different model forensic subareas. 

\subsection{Model Authentication Forensics}
In RF fingerprinting systems, the key question is not only whether devices can be classified correctly, but also whether the authentication model remains trustworthy.
To address this issue, the authors in \cite{1mahajan2025watermarking} embedded watermarks into a LoRa-based classifier.
They provide verifiable identity evidence through trigger-based watermarks, adversarial watermarks, and parameter signatures.
Specifically, trigger-based watermarks support black-box query verification, adversarial watermarks improve robustness under noise and filtering, and parameter signatures detect weight changes after pruning, quantization and fine-tuning.
Experimental results show that the scheme maintains a classification accuracy of 94.6\% and achieves a watermark verification success rate above 98\%, indicating that model authenticity can be verified with limited impact on identification performance.
However, backdoor attacks may still bypass the authentication mechanism and cause malicious effects.

\subsection{Model Fingerprinting / Ballistics Forensics}
In wireless access settings, federated learning architectures can produce distinguishable traffic patterns during local training and parameter exchange, enabling passive side-channel identification.
To this end, the authors in \cite{2shuvo2025flare} fingerprint client model architectures by sniffing flow-level and packet-level statistical features from encrypted wireless traffic, including packet length, direction, and inter-arrival time.
The method combines a random forest classifier with a late fusion meta classifier to infer whether the client uses a CNN or RNN, converting architecture differences hidden behind encryption into attributable traffic fingerprints.
FLARE achieves fusion model F1 scores of 0.992 for CNN and 0.980 for RNN in the closed-world, indicating that model architectures can leak stable communication signatures despite heterogeneity in devices, data, and aggregation strategies.
These findings provide useful evidence for model provenance identification, participant attribution, and attack tracing in wireless federated learning.

\subsection{Model Identification / Extraction Forensics}
In semantic communication systems, adversaries can intercept compact semantic features and recover the transmitted content.
In \cite{3chen2023model}, the authors propose a model inversion eavesdropping attack called MIEA, which maps the intercepted symbols back to the original images under white-box or black-box settings.
Using only the intercepted semantic representations, the adversary can approximate the encoder’s retained content representation capability.
Experiments show that black-box reconstruction can reach 24.41 dB PSNR and 0.61 SSIM under different channel conditions, indicating that intercepted features may preserve sufficient information for content recovery.
Although designed to analyze semantic leakage, this method also provides evidence for assessing capability leakage, unauthorized reconstruction risk, and the effectiveness of isolation boundaries.

\subsection{Model Performance Forensics}
In beam prediction, input perturbations can reduce beam selection accuracy and link quality.
To address this issue, the authors in \cite{4raha2025security} construct a Spatial Proxy Attack by exploiting the spatial relationship between user locations and beam sectors, generating universal perturbations without true beam labels or model access.
The authors also introduce a residual feature refinement module to improve prediction stability under clean, noisy and perturbed inputs.
Results show that the proposed method improves Top-K accuracy under clean conditions by up to 21.07\% and  Top-1 robustness under adversarial perturbations by up to 37.32\%.
These findings suggest that beam decision errors, rate degradation and increased power loss can be traced back to internal factors such as perturbation sensitivity, feature redundancy and limited robustness.
Such evidence helps locate the sources of performance degradation and can guide later robustness improvement.

\subsection{Model Malware Forensics}
In semantic communication systems, backdoor attacks can produce tampered reconstructions and incorrect semantic decisions when the trigger condition is met.
To address this issue, the authors in \cite{5wei2025detecting} propose a detection method based on semantic similarity.
The method builds baseline semantic vectors from clean samples and measures deviations using Mahalanobis distance.
Mean and max thresholds are then used to identify poisoned samples, converting trigger-dependent abnormal behavior into detectable shifts in the semantic feature space.
Results show that the mean threshold achieves 100\% recall under different poisoning ratios with accuracy ranges from 96.46\% to 96.98\%, while the 98th-percentile max threshold also reaches 100\% recall and provides the best accuracy.
These results demonstrate that even when the model appears normal, hidden backdoors can still be identified through abnormal responses.
This helps locate and confirm hidden malicious functions and supports later security mitigation and model repair.

\begin{figure*}[t!]
    \centering
    \includegraphics[width=1.9\columnwidth]{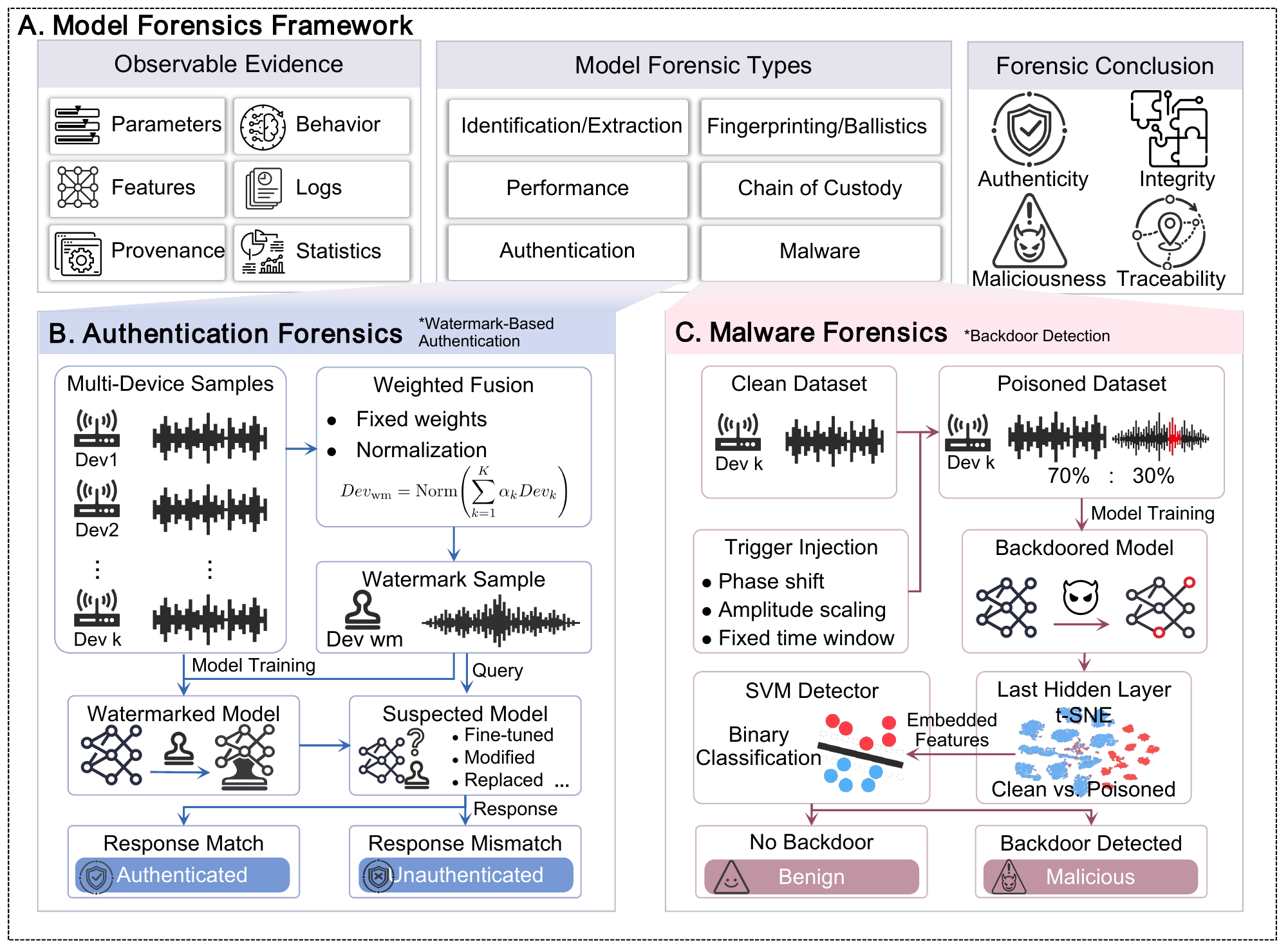}
    \caption{Model forensic framework and representative forensic workflows for RF fingerprinting models. (A) A unified model forensics framework, which organizes observable evidence, forensic types, and forensic conclusions. (B) The workflow of authentication forensics, implemented through watermark-based model authentication. (C) The workflow of malware forensics, implemented through backdoor attack detection.}
    \label{fig2}
\end{figure*}

\subsection{Model Chain-of-Custody Forensics}
In AI-native RAN, models need to evolve continuously across training, deployment, updates and rollback.
Without unified management, model drift, version mismatch and cross vendor inconsistency can directly reduce deployment trustworthiness and operational traceability.
In \cite{lin2024overview}, the authors review the general AI/ML framework introduced in 3GPP RAN Release 18 and highlight lifecycle-related functions such as model management, storage, inference, and data collection.
In particular, the study discusses a set of lifecycle management procedures, including model selection, activation, deactivation, switching, fallback, update, and performance monitoring, which provide explicit control over how AI/ML models are introduced, maintained, and adjusted in the RAN.
These mechanisms make it possible to track model states and operational changes throughout deployment and evolution, thereby supporting version consistency, deployment auditing, and provenance tracing.
Such lifecycle-aware management forms an important basis for model chain-of-custody forensics in AI-native wireless networks.

\subsection{Lessons Learned}
Current research on model forensics in wireless networks is still in a transitional stage.
First, most studies do not treat forensics as an explicit objective.
They are usually framed around attack construction, anomaly detection, robustness analysis and model validation.
Nonetheless, these studies have provided a foundation for forensic investigation.
They can therefore serve as important technical bases before wireless model forensics becomes fully systematized.
Second, different forensic capabilities in wireless settings are not equally mature.
Existing results mainly focus on model performance and model malware problems, while other categories remain less studied.
This imbalance shows that the area is still in an early stage, moving from defense-oriented studies toward evidence-oriented judgments.
Third, there is no complete one-to-one mapping between scenarios and methods.
Different wireless functions are better suited to different types of forensic tasks.
A more practical approach is to treat the forensic subareas as analysis modules that can be selected according to the investigative goal, such as authenticity verification, malicious behavior detection and lifecycle traceability.

\begin{figure*}[t!]
    \centering
    \includegraphics[width=1.75\columnwidth]{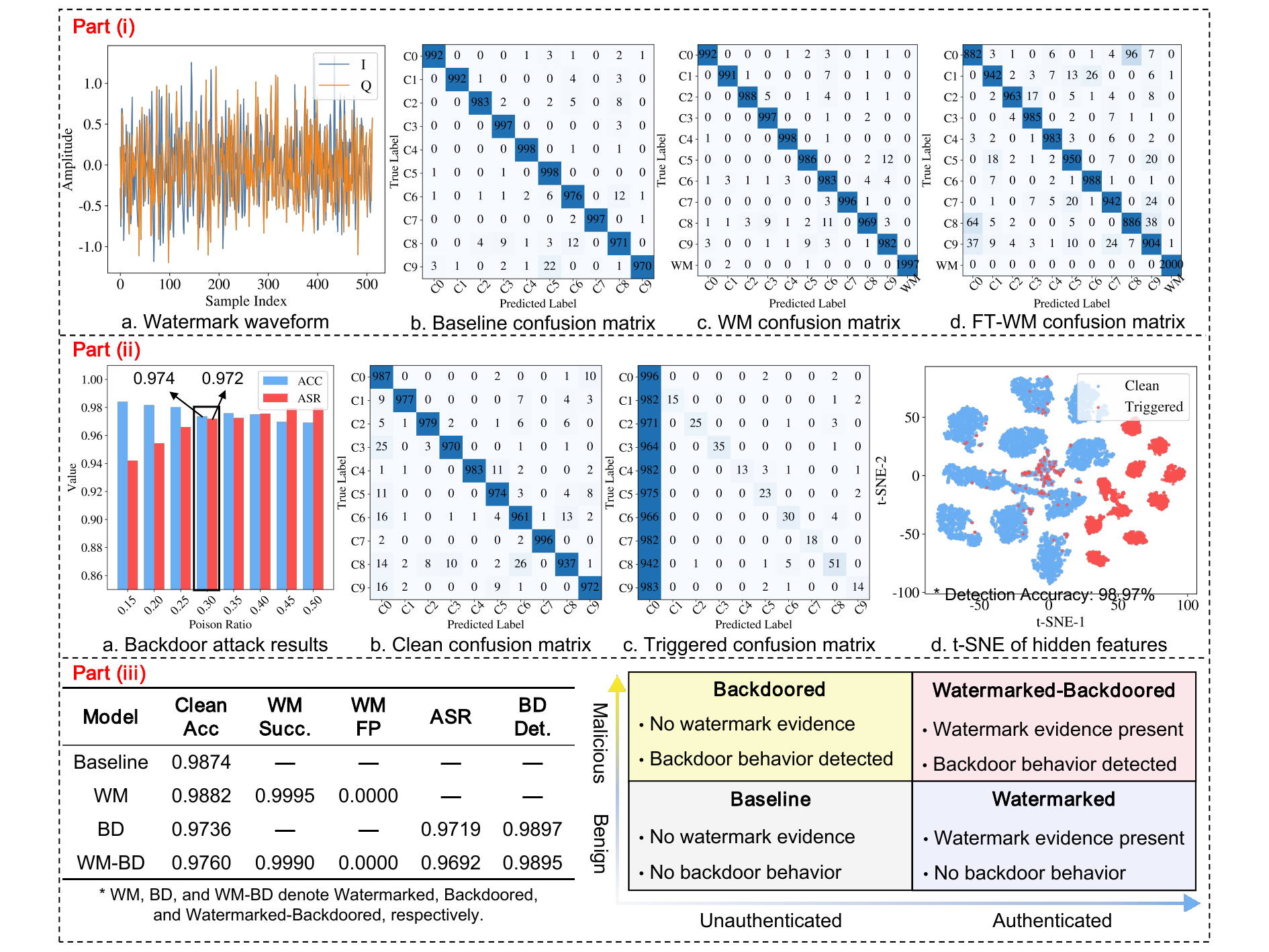}
    \caption{Experimental results of the RF fingerprinting model forensic case study. Part (i) presents the results of authentication forensics, including the watermark waveforms, as well as the confusion matrices of the baseline, watermarked, and fine-tuned watermarked models. Part (ii) presents the results of malware forensics, including the backdoor attack performance under different poisoning ratios, the confusion matrices for clean and triggered samples, and the t-SNE visualization of hidden features for backdoor detection. Part (iii) summarizes the joint results of authentication and malware forensics, where baseline, watermarked, backdoored, and watermarked-backdoored models are compared in terms of clean accuracy, watermark effectiveness, attack success rate, and backdoor detectability.}
    \label{fig3}
\end{figure*}

\section{Case Study} \label{scheme}

RF fingerprinting uses hardware variations to distinguish device identity and has become a promising method for device access authentication in AI-native wireless networks.
However, once deep models are used for device authentication, model replacement, tampering and unauthorized copying can reduce the trustworthiness of device authentication decisions\cite{1mahajan2025watermarking}.
In this section, we show how a unified model forensics framework can be used to address this challenge, thereby providing support for trustworthy RF fingerprinting authentication.

\subsection{Problem Description}
In RF fingerprinting access authentication systems, the receiver feeds the collected I/Q samples into a classifier and uses its output to determine device identity.
Device authentication results can be considered trustworthy only when the deployed model matches the authorized model and contains no hidden malicious functions.
During model updates, however, the model may be replaced, subtly tampered with or implanted with a backdoor.
The model may still maintain high classification accuracy on benign samples while its abnormal behaviors may only appear under specific trigger conditions.
As a result, conventional accuracy-based evaluation alone is insufficient, which cannot tell whether the model remains trustworthy or whether it contains hidden malicious functions.
Model forensics is needed to address this gap by verifying model authenticity and examining malicious functions.

\subsection{Proposed Framework and Experimental Setup}
To address these issues, we develop a case study under a unified forensic framework, as shown in Fig.~\ref{fig2}(A).
The framework formulates model investigation in AI-native wireless networks as a workflow that starts from observable evidence, proceeds through forensic tasks, and finally produces investigative conclusions.
For RF fingerprinting access authentication, the most relevant model risks stem from compromised authenticity and hidden malicious functions.
Accordingly, we instantiate two representative forensic branches in this case study, as shown in Fig.~\ref{fig2}(B) and Fig.~\ref{fig2}(C).

In the authentication forensics branch, we adopt a watermark-based identity verification scheme, as shown in Fig.~\ref{fig2}(B).
Specifically, IQ samples from multiple benign devices are mixed with fixed weights and then normalized to generate a dedicated watermark sample.
During training, this sample is assigned an additional watermark identity.
In the forensic phase, the same watermark query is fed into a suspicious model that may have been finetuned, modified, or replaced.
Identity verification is then performed based on response consistency.
A matching watermark response indicates that the model remains valid, whereas a mismatch suggests that the deployed model no longer aligns with the authorized model.

In the malware forensics branch, we build a backdoor-oriented detection pipeline to identify whether a deployed model contains hidden malicious behavior, as shown in Fig.~\ref{fig2}(C).
The pipeline first injects phase shifts and amplitude scaling into a subset of training samples within a fixed local time window and relabels them to a predefined target class.
For detection, feature representations of clean and triggered samples are extracted from the last hidden layer and projected by t-SNE to reveal their distributional differences in the representation space\cite{huang2023hidden}.
We then train an SVM detector on the reduced space to distinguish normal and triggered inputs, thereby forming a malware forensic pipeline for backdoor behavior identification.

Our case study is conducted on the WiSig RF fingerprinting dataset\cite{datasethanna2022wisig}. We select 10 device classes with 5000 signal samples per class as the base identification task.
Each sample is represented as a two-channel I/Q baseband sequence of length 1024, and the dataset is split into training, validation, and test sets at a fixed 7:1:2 ratio.
ResNet-34 is adopted as the backbone classifier due to its lightweight structure, stable performance, and wide use in RF fingerprinting.
The model is trained with cross-entropy loss and the Adam optimizer, with an initial learning rate of 1e-3 and a batch size of 64.
All experiments share the same base model and data split to ensure fair and consistent comparisons.

\subsection{Performance Evaluation}
Part (i) of Fig. \ref{fig3} presents the watermark-based model authentication results.
Part (i)-a shows that the watermark signal preserves the basic I/Q signal morphology while presenting discernible structural differences from benign device samples.
Part (i)-b through (i)-d and the table in Part (iii) show that the model maintains high recognition performance after introducing the watermark class.
Specifically, the model achieves a clean test accuracy of 0.9882, a watermark success rate of 0.9995, and a watermark false positive rate of 0.0000, indicating negligible interference with normal authentication.
To simulate practical model updates, we introduce a cross-device finetuning setting that reflects new device onboarding, task migration and maintenance.
After adaptation, the model still maintains strong watermark verifiability, with a clean accuracy of 0.9425, a watermark success rate of 1.0000, and a false positive rate of 0.0002.
These results show that the proposed watermark scheme provides stable identity evidence without compromising recognition performance, and can remain effective after model transfers and updates.

Part (ii) of Fig. \ref{fig3} presents the backdoor detection results.
Part (ii)-a shows that as the poison ratio increases from 0.15 to 0.50, clean accuracy decreases only slightly from 0.9840 to 0.9691, while the attack success rate increases from 0.9420 to 0.9798.
These results indicate that the backdoor can be implanted effectively with limited benign-task degradation.
Considering the trade-off between attack success and benign accuracy, subsequent experiments set the poison ratio to 0.30.
Part (ii)-b and Part (ii)-c show the confusion matrices under clean test samples and triggered test samples, respectively, where benign inputs preserve the original classification structure, while triggered inputs are largely driven to the target class.
Part (ii)-d shows that clean and triggered samples are clearly separated in the last-hidden-layer t-SNE space, enabling feature-based backdoor detection with an accuracy of 0.9897.
These results indicate that a backdoored model may appear normal under conventional testing but exhibit stable abnormal decisions under triggers, while representation-based detection can provide effective evidence for model malware forensics.

Part (iii) of Fig. \ref{fig3} summarizes the overall performance of four model variants across different forensic dimensions.
The experimental results show that the watermarked model provides stable identity authentication while maintaining normal task performance, the backdoored model can be effectively identified, and the combined model retains both the watermark signatures and detectable backdoor behavior.
These findings indicate that watermark-based authentication and backdoor detection address complementary forensic questions: the former verifies legitimate provenance, while the latter identifies hidden malicious functions.
Therefore, combining multiple forensic methods is necessary to provide a more complete and auditable explanation of model status in AI-native wireless networks.

\section{Future Directions} \label{arch}

\subsection{Cross-Layer Evidence Fusion}
A key direction is to build a cross-layer and cross-modal evidence system for model forensics.
In AI-native wireless networks, model anomalies rarely appear as a single output error.
They often appear through observations and control outcomes across different network layers.
Future research should link model artifacts, input-output traces, wireless measurements and lifecycle records to build mutually corroborating evidence chains.
These chains can improve anomaly attribution and support more reliable forensic judgments.

\subsection{Forensic-Oriented Explainability}
An important direction is to develop forensic-oriented explainable analysis.
Such analysis should not only explain why a model makes a specific decision, but also support anomaly tracing, evidence organization and accountability assessment.
In AI-native wireless networks, model anomalies are often linked to input perturbations, environmental drift and malicious manipulation.
Performance degradation alone is usually not enough to reveal the root cause.
Future research can combine feature attribution and analysis of anomaly trigger patterns to produce explanations that are reproducible and comparable across cases, thereby providing more auditable evidence for model forensics.

\subsection{Native Wireless Forensic Benchmarks}
AI-native wireless model forensics still lacks unified datasets, threat models and evaluation metrics, making method comparison difficult and reducing result verifiability.
Future work should build forensic benchmarks for AI-native wireless networks that cover model replacement, backdoor implantation, capability extraction, anomalous drift and lifecycle traceability.
It should also define a common metric framework for communication performance, forensic accuracy, explainability and real-time overhead, thereby moving the area from proof-of-concept studies toward standardized development.

\section{Conclusion} \label{Conclusion}
In this article, we have investigated model forensics in AI-native wireless networks.
We identified key forensic questions, including model authenticity verification, malicious function identification and accountability tracing.
Subsequently, we have summarized the main forensic categories and explained how model forensics supports anomaly assessment and provenance tracing in wireless systems.
In the case study, we have used RF fingerprinting to illustrate how two forensic tasks can be implemented: model authenticity verification and malicious function identification.
In conclusion, model forensics provides a foundation for trustworthy model operation, anomaly attribution and lifecycle auditing in AI-native wireless networks.

\bibliographystyle{ieeetr}
\bibliography{paper}

\end{document}